\documentclass[12pt]{article}
\usepackage{overcite}
\usepackage{amsfonts}

\title{ {\bf Superintegrable systems with spin in two- and three-dimensional Euclidean spaces}}
\author{\vspace{1cm}\\
         {\bf Pavel Winternitz}
         \thanks{E-mail address:
        wintern@crm.umontreal.ca}
         {\,\,and \bf \.{I}smet Yurdu\c{s}en}
        \thanks{E-mail address:
       yurdusen@crm.umontreal.ca} \\Centre de Recherches Math\'{e}matiques, Universit\'{e} de Montr\'{e}al,\\ CP 6128, Succ. Centre-Ville, Montr\'{e}al, Quebec H3C 3J7, Canada}
\date{\today}
\begin{document}
\setlength{\baselineskip}{24pt} 
\maketitle
\setlength{\baselineskip}{5mm}
\begin{abstract}
The concept of superintegrability in quantum mechanics is extended to the
case of a particle with spin $s=1/2$ interacting with one of spin $s=0$. 
Non-trivial superintegrable systems with $8$- and $9$-dimensional Lie 
algebras of first-order integrals of motion are constructed in two- and 
three-dimensional spaces, respectively.
\end{abstract}


\section{Introduction}\label{pwiy:sec1}
A superintegrable system in classical and quantum mechanics is a system with 
more integrals of motion than degrees of freedom. A large body of literature 
on such systems exists and is mainly devoted to quadratic 
superintegrability. This is the case of a scalar particle in a potential
$V(\vec{r})$ in an  $n$-dimensional space with $k$ integrals of motion, 
$n+1\le k \le 2n-1$, all of them first- or second-order polynomials in
the momenta (see e.g. 
\cite{Winternitz, Makarov, Rodriguez, Kalnins, Kalnins2, Tempesta} 
and references therein). Maximally superintegrable systems have 
$2n-1$ integrals of motion and are of
special interest. In classical mechanics all bounded trajectories in such 
systems are closed. In quantum mechanics these systems have degenerate
energy levels and it has been conjectured that they are all exactly 
solvable \cite{Tempesta}.

Quadratic integrability for a Hamiltonian of the form
\begin{equation}
H=\frac{1}{2} \vec{p}^{\,2} + V(\vec{r})\,, \label{pwiy:eq1}
\end{equation}
i.e. the existence of $n$ second-order integrals of motion in involution, 
is related to the separation of variables in the Hamilton--Jacobi, or the 
Schr\"{o}dinger equation, respectively. Quadratically superintegrable systems
are multiseparable. The non-abelian algebra of integrals of motion usually 
has several non-equivalent $n$-dimensional Abelian subalgebras, each of them
leading to the separation of variables in a different coordinate system.

The situation changes when one goes beyond Hamiltonians of the 
type of (\ref{pwiy:eq1}), or considers higher-order integrals of motion. If
a vector potential is added in (\ref{pwiy:eq1}), corresponding to velocity
dependent forces, e.g. a magnetic field, then second-order integrability no
longer implies the separation of variables 
\cite{Dorizzi, Charest, Pucacco, Benenti} and the same is true in the case
of third-order integrals of motion 
for (\ref{pwiy:eq1}) \cite{Gravel, Gravel2, Marquette}.

The purpose of this contribution is to report on a research program which
investigates the concepts of integrability and superintegrability
for systems involving particles with spin. 

Here we restrict ourselves to the simplest case of the interaction of 
two particles with spin $0$ and spin $1/2$, respectively. We write the 
Schr\"{o}dinger--Pauli equation including a spin--orbit term as 
\begin{eqnarray}
H\Psi=\left[-\frac{1}{2} \Delta + V_0(\vec{r}) 
+ \frac{1}{2}\Big\{V_1(\vec{r}), \,\vec{\sigma} \!\cdot \! 
\vec{L}\Big\}\right]\Psi
\,, \label{pwiy:eq2}
\end{eqnarray}
where $\{,\}$ denotes an anticommutator, $\sigma_1$, $\sigma_2$, 
$\sigma_3$ are the usual Pauli matrices, $\Psi$ is a two-component
spinor and $L$ is the angular momentum operator. The Hamiltonian
given in (\ref{pwiy:eq2}) would describe, for instance a low energy 
(nonrelativistic) pion--nucleon interaction. In this paper we 
restrict ourselves to first-order integrability. Thus we require 
that the integrals of motion should be first-order matrix 
differential operators 
\begin{eqnarray}
X=\frac{1}{2}\sum_{\mu=0}^3 \sum_{k=1}^3
\left[A_{\mu k}(\vec{r})\sigma_{\mu}p_{k}+
\sigma_{\mu}p_{k} A_{\mu k}(\vec{r})\right]+
\sum_{\mu=0}^3 \phi_{\mu}(\vec{r})\sigma_{\mu}
\,, \label{pwiy:eq3}
\end{eqnarray}
with $\sigma_0\equiv I_2$. For particles with spin zero 
only components with $\mu=0$ in (\ref{pwiy:eq3}) would survive and 
the condition $[H,X]=0$ (with $V_1=0$), would imply a simple geometric
symmetry.

\section{The Two-Dimensional Case}\label{pwiy:sec2}
Let us first consider the case when motion is constrained to a
Euclidean plane. We assume $\Psi(\vec{r})=\Psi(x,y)$, set
$p_3=0$, $z=0$ and write the Schr\"{o}dinger--Pauli equation given
in (\ref{pwiy:eq2}) as 
\begin{equation}
H \Psi=\left[\frac{1}{2}({p_1}^2+{p_2}^2)+V_0(x,y)+V_1(x,y){\sigma}_3 L_3 +
\frac{1}{2}{\sigma}_3(L_3 V_1(x,y)) \right] \Psi 
\label{pwiy:eq4}
\end{equation}
with 
\begin{eqnarray}
p_1=-i\partial_x, \qquad p_2=-i\partial_y, \qquad 
L_3=i(y\partial_x-x\partial_y) 
\,.\nonumber
\end{eqnarray}
The operator (\ref{pwiy:eq3}) reduces to 
\begin{eqnarray}
X&=&(A_0 p_1 + B_0 p_2 + \phi_0) I + (A_1 p_1 +B_1 p_2 + \phi_1)\sigma_3 
\nonumber \\
&\,& +\frac{1}{2}\left[ \big((p_1 A_0)+(p_2 B_0)\big) I + 
\big((p_1 A_1)+(p_2 B_1) \big) \sigma_3 \right]
\,. \label{pwiy:eq5}
\end{eqnarray}
The commutativity condition $[H,X]=0$ implies $12$ determining equations 
for the $8$ functions $A_{\mu}(x,y)$, $B_{\mu}(x,y)$, $\phi_{\mu}(x,y)$ 
and $V_{\mu}(x,y)$ ($\mu=0,1$). From these we obtain
\begin{eqnarray}
A_{\mu}=\omega_{\mu}y+a_{\mu}, \qquad B_{\mu} = -\omega_{\mu}x + b_{\mu}
\,, \nonumber
\end{eqnarray}
\begin{eqnarray}
\phi_{\mu, x}&=& \delta_{\mu, 1-\nu}[-b_{\nu}V_1-(\omega_{\nu}y+a_{\nu})y 
V_{1,x}+(\omega_{\nu}x-b_{\nu})y V_{1,y}]\,, \nonumber \\
\phi_{\mu, y}&=& \delta_{\mu, 1-\nu}[a_{\nu}V_1+(\omega_{\nu}y+a_{\nu})x 
V_{1,x}-(\omega_{\nu}x-b_{\nu})x V_{1,y}]\,, \nonumber
\end{eqnarray}
\begin{eqnarray}
(\omega_{\mu}y+a_{\mu})V_{0,x}+(-\omega_{\mu}x+b_{\mu})V_{0,y}=
\delta_{\mu, 1-\nu}(x\phi_{\nu,y}-y\phi_{\nu,x})V_1\,,
\label{pwiy:eq6}
\end{eqnarray}
where $\omega_{\mu}$, $a_{\mu}$ and $b_{\mu}$ are real constants 
and ($\mu, \nu$)$=$($0,1$). The above equations can be simplified by 
rotations in the $xy$-plane and by gauge transformations of 
the form
\begin{equation}
\widetilde {H_{\,}}=U^{-1}HU, \qquad U=\left(\begin{array}{cc} 
e^{i\alpha}& 0\\
0& e^{-i\alpha} \end{array}\right), \qquad \alpha=\alpha(\xi), 
\qquad \xi=\frac{y}{x}
\,. \label{pwiy:eq7}
\end{equation}
The gauge transformations leave the kinetic energy invariant but modify
the potentials
\begin{eqnarray}
\widetilde {V_1}=V_1 + \frac{\dot {\alpha}}{x^2}, \qquad 
\widetilde {V_0}=V_0+(1+\frac{y^2}{x^2})(\frac{1}{2}\frac{{\dot {\alpha}}^2}{x^2}
+ \dot {\alpha}V_1)
\,. \label{pwiy:eq8}
\end{eqnarray}

The results obtained by analyzing (\ref{pwiy:eq6}) can be summed up as 
follows:
\begin{enumerate}
\item Exactly one superintegrable system with $V_1\neq0$ exists up to 
gauge transformation, namely
\begin{equation}
H=-\frac{1}{2}\Delta + \frac{1}{2}{\gamma}^2(x^2+y^2) + \gamma \sigma_3 L_3,
\qquad \gamma=\mathrm{const}
\,. \label{pwiy:eq9}
\end{equation}
It allows an $8$-dimensional Lie algebra $\mathcal{L}$ of first-order 
integrals of motion with a basis given by 
\begin{eqnarray}
L_{\pm}&=&i(y\partial_x-x\partial_y)(I\pm \sigma_3)\,,\nonumber \\
X_{\pm}&=&(i\partial_x \mp \gamma y )(I\pm \sigma_3)\,,\nonumber \\
Y_{\pm}&=&(i\partial_y \pm \gamma x )(I\pm \sigma_3)\,,\nonumber \\
I_{\pm}&=& I\pm \sigma_3
\,. \label{pwiy:eq10}
\end{eqnarray}
The algebra $\mathcal{L}$ is isomorphic to the direct sum of two central 
extensions of the Euclidean Lie algebra $e(2)$
\begin{eqnarray}
\mathcal{L} \sim \widetilde {e}_+(2)\oplus\widetilde {e}_-(2)\,, \qquad
\widetilde {e}_{\pm}(2)=\{L_{\pm}, X_{\pm}, Y_{\pm}, I_{\pm}\}
\,.\label{pwiy:eq11}
\end{eqnarray}
The two Casimir operators of $\mathcal{L}$ and the Hamiltonian (\ref{pwiy:eq9}) are
\begin{eqnarray}
C_{\pm}=X_{\pm}^2 + Y_{\pm}^2 \pm 4\gamma L_{\pm}I_{\pm}\,,
\qquad H=\frac{1}{8}\left(C_{+}+C_{-}\right)
\,. \label{pwiy:eq12}
\end{eqnarray}
Conjugacy classes of elements of the algebra $\mathcal{L}$ can be represented by 
\begin{eqnarray}
X_1=L_+ + \lambda L_-, \,\, X_2= L_+ + \lambda X_-, \,\, X_3=X_+ + 
\lambda X_-\,,\,\, \lambda \in \mathbb R
\,.\label{pwiy:eq13}
\end{eqnarray}
\item Integrable systems (with one integral of motion in addition to $H$)
exist. They are given by
\begin{enumerate}
\item 
\begin{eqnarray}
V_0=V_0(\rho)\,, \qquad V_1=V_1(\rho)\,, \qquad \rho=\sqrt{x^2+y^2} 
\,, \nonumber \\
X =(\omega_0 + \omega_1 \sigma_3)L_3\,, \qquad \omega_{\mu}=\mathrm{const}
\,,\qquad \mu=0,1
\,. \label{pwiy:eq14}
\end{eqnarray}
\item 
\begin{eqnarray}
V_1=V_1(x)\,, &\,& \qquad V_0=\frac{y^2}{2} {V_1}^2(x) + F(x)\,, \nonumber \\
X&=&-i\partial_y - \sigma_3 \int V_1(x)dx
\,. \label{pwiy:eq15}
\end{eqnarray}
\end{enumerate}
\end{enumerate}
Thus, the superintegrable system (\ref{pwiy:eq9}) involves one arbitrary 
constant $\gamma$. The integrable systems (\ref{pwiy:eq14}) and 
(\ref{pwiy:eq15}) each involve two arbitrary functions of
one variable.

The integrals of motion can be used to solve the  Schr\"{o}dinger--Pauli 
equation for the superintegrable system (in several different manners). In
the two integrable cases (\ref{pwiy:eq14}) and 
(\ref{pwiy:eq15}) they can be used to reduce the problem to solving
ordinary differential equations. For all details see the original 
article \cite{Winternitz2}.

Before going over to the case $n=3$ let us mention that two important 
features that simplify the case $n=2$. The first one is that the Hamiltonian 
given in (\ref{pwiy:eq4}) is a diagonal matrix operator (since $\sigma_2$ and
$\sigma_3$ do not figure). Hence we could restrict our search to integrals 
$X$ that are also diagonal. The second is that there exists a zeroth-order 
integral $X=\sigma_3$ (for any $V_0$ and $V_1$), in addition to the 
trivial commuting operator $X=I$. Hence any integral of motion can be 
multiplied by $\sigma_3$ and there is a ``doubling'' of the number of 
integrals of a given order.

We have set the Planck constant $\hbar=1$ in all calculations. Keeping
$\hbar$ in the Hamiltonian and integrals of motion does not change any
of the conclusions. In particular $V_0$ and $V_1$ do not depend on $\hbar$.

\section{The Three-Dimensional Case}\label{pwiy:sec3}
Let us now consider (\ref{pwiy:eq2}) and search for an 
integral of the form (\ref{pwiy:eq3}) which we rewrite as
\begin{eqnarray}
X=(A_0+\vec{A} \cdot \vec{\sigma})p_1+(B_0+\vec{B} \cdot \vec{\sigma})p_2+
(C_0+\vec{C} \cdot \vec{\sigma})p_3+\phi_0+\vec{\phi}\,\cdot \vec{\sigma} 
\nonumber \\
-\frac{i}{2}\left\{(A_0+\vec{A} \cdot \vec{\sigma})_x
+(B_0+\vec{B} \cdot \vec{\sigma})_y+(C_0+\vec{C} \cdot \vec{\sigma})_z
\right\}
\,, \label{pwiy:eq16}
\end{eqnarray}
where $A_0$, $B_0$, $C_0$, $\phi_0$ and $A_i$, $B_i$, $C_i$, $\phi_i$ 
($i=1,2,3$) are all functions of $\vec{r}$, to be determined from the 
commutativity condition $[H,X]=0$. This commutator will have second-, 
first- and zeroth-order terms in the momenta. 

From the second-order terms we obtain
\begin{equation}
A_0=b_1-a_3 y+a_2 z ,\,\, B_0=b_2+a_3 x-a_1 z , \,\,
C_0=b_3-a_2 x+a_1 y 
\,, \label{pwiy:eq17}
\end{equation}
where $a_i$ and $b_i$ are constants. We also obtain the following 
overdetermined system of $18$ first-order quasilinear partial differential 
equations (PDE) for $A_i$, $B_i$, $C_i$ and $V_1$
\begin{eqnarray}
&2zA_1V_1+A_{3,x}=0\,, \quad 2yA_1V_1+A_{2,x}=0\,, \quad 2xB_2V_1+B_{1,y}=0
\,, \nonumber \\
&2zB_2V_1+B_{3,y}=0\,, \quad 2xC_3V_1+C_{1,z}=0\,, \quad 2yC_3V_1+C_{2,z}=0
\,, \nonumber \\
&2V_1\big(yA_2+zA_3\big)-A_{1,x}=0\,, \quad 2V_1\big(xB_1+zB_3\big)-B_{2,y}=0
\,, \nonumber \\
&2V_1\big(xC_1+yC_2\big)-C_{3,z}=0\,, \quad 2zV_1\big(A_2+B_1\big)+A_{3,y}
+B_{3,x}=0\,, \nonumber  \\
&2yV_1\big(A_3+C_1\big)+A_{2,z}+C_{2,x}=0\,, \quad 2xV_1\big(B_3+C_2\big)
+B_{1,z}+C_{1,y}=0\,, \nonumber \\
&2V_1\big(xA_1+yA_2-zC_1\big)-A_{3,z}-C_{3,x}=0\,, \nonumber  \\
&2V_1\big(xB_1+yB_2-zC_2\big)-B_{3,z}-C_{3,y}=0\,, \nonumber  \\
&2V_1\big(xA_2-yB_2-zB_3\big)+A_{1,y}+B_{1,x}=0\,, \nonumber \\
&2V_1\big(xA_1+zA_3-yB_1\big)-A_{2,y}-B_{2,x}=0\,, \nonumber \\
&2V_1\big(xA_3-yC_2-zC_3\big)+A_{1,z}+C_{1,x}=0\,, \nonumber \\
&2V_1\big(yB_3-xC_1-zC_3\big)+B_{2,z}+C_{2,y}=0
\,. \label{pwiy:neweq1}
\end{eqnarray}

For any $V_1$ (\ref{pwiy:neweq1}) has the following solution
\begin{eqnarray}
A_1=0, \qquad A_2=zw, \qquad A_3=-yw, \nonumber \\
B_1=-zw, \qquad B_2=0, \qquad B_3=xw, \nonumber \\
C_1=yw, \qquad C_2=-xw, \qquad C_3=0, 
\label{pwiy:neweq2}
\end{eqnarray}
where $w$ is an integration constant.

The first-order terms provide a system of
$9$ first-order quasilinear PDE for $V_1$ and $\phi_i$ and
$3$ first-order quasilinear PDE for $\phi_0$ and
$A_i$, $B_i$, $C_i$. They also provide $9$ second-order PDE 
for $A_i$, $B_i$, $C_i$ and $V_1$, however, these are differential 
consequences of (\ref{pwiy:neweq1}). The $12$ first-order quasilinear PDE
can be written as
\begin{eqnarray}
&V_1(b_1-a_3 y +2 y \phi_3)+x(A_0 V_{1x}+B_0 V_{1y}+C_0 V_{1z})+\phi_{2z}=0\,, \nonumber \\
&V_1(b_1+a_2 z -2 z \phi_2)+x(A_0 V_{1x}+B_0 V_{1y}+C_0 V_{1z})-\phi_{3y}=0\,, \nonumber \\
&V_1(b_2-a_1 z +2 z \phi_1)+y(A_0 V_{1x}+B_0 V_{1y}+C_0 V_{1z})+\phi_{3x}=0\,, \nonumber \\
&V_1(b_2+a_3 x -2 x \phi_3)+y(A_0 V_{1x}+B_0 V_{1y}+C_0 V_{1z})-\phi_{1z}=0\,, \nonumber \\
&V_1(b_3-a_2 x +2 x \phi_2)+z(A_0 V_{1x}+B_0 V_{1y}+C_0 V_{1z})+\phi_{1y}=0\,, \nonumber \\
&V_1(b_3+a_1 y -2 y \phi_1)+z(A_0 V_{1x}+B_0 V_{1y}+C_0 V_{1z})-\phi_{2x}=0\,, \nonumber \\
&V_1(a_2 y+a_3 z -2 y \phi_2- 2 z \phi_3)+\phi_{1x}=0\,, \nonumber \\
&V_1(a_1 x+a_3 z -2 x \phi_1- 2 z \phi_3)+\phi_{2y}=0\,, \nonumber \\
&V_1(a_1 x+a_2 y -2 x \phi_1- 2 y \phi_2)+\phi_{3z}=0
\,, \label{pwiy:neweq3}
\end{eqnarray}
where $A_0$, $B_0$ and $C_0$ are given in (\ref{pwiy:eq17}) and 
\begin{eqnarray}
\phi_{0x}&=&V_1\Big((yA_{3x}-xA_{3y})+(xA_{2z}-zA_{2x})+(zA_{1y}-yA_{1z})+(C_2-B_3)\Big) \nonumber \\
&\,&+V_{1x}(zA_2-yA_3)+V_{1y}(zB_2-yB_3)+V_{1z}(zC_2-yC_3)\,, \nonumber \\
\phi_{0y}&=&V_1\Big((yB_{3x}-xB_{3y})+(xB_{2z}-zB_{2x})+(zB_{1y}-yB_{1z})+(A_3-C_1)\Big) \nonumber \\
&\,&+V_{1x}(xA_3-zA_1)+V_{1y}(xB_3-zB_1)+V_{1z}(xC_3-zC_1)\,, \nonumber  \\
\phi_{0z}&=& V_1\Big((yC_{3x}-xC_{3y})+(xC_{2z}-zC_{2x})+(zC_{1y}-yC_{1z})+(B_1-A_2)\Big) \nonumber \\
&\,&+V_{1x}(yA_1-xA_2)+V_{1y}(yB_1-xB_2)+V_{1z}(yC_1-xC_2)
\,. \label{pwiy:neweq4} 
\end{eqnarray}

The system of $9$ PDE given in (\ref{pwiy:neweq3}) has a solution
if the following conditions are satisfied:
\begin{enumerate}
\item If $b_i\neq0$, $i=1,2,3$, then $V_1=\frac{1}{r^2}$.
\item If $b_i=0$, $\forall i$, then $V_1=V_1(r)$.
\end{enumerate}

Finally, the zeroth-order terms in the commutator 
provide $8$ more PDE
that also involve $V_0$ and are in general of second-order. 
In fact some of them are third-order differential equations,
however, by using (\ref{pwiy:neweq1}) they can be reduced the
second-order ones. These equations are too long to be presented 
here.

The complete discusion of the above determining equations is long and 
we cannot reproduce the details here so we just present some results.

\noindent {\bf (a) A superintegrable system.}

The entire overdetermined system of equations can be solved for 
$V_0\!=\!\frac{1}{r^2}$, $V_1=\frac{1}{r^2}$.
We obtain the Hamiltonian
\begin{equation}
H=-\frac{1}{2}\Delta + \frac{1}{r^2} + 
\frac{1}{r^2} (\vec{\sigma}\,,\vec{L})
\,, \label{pwiy:eq18}
\end{equation}
with a $9$-dimensional Lie algebra $\mathcal{L}$ of integrals of
motion:
\begin{eqnarray}
J_i=L_i+\frac{1}{2}\sigma_i\,,  &\,& \qquad
\Pi_i=p_i-\frac{1}{r^2}\epsilon_{ikl}x_k\sigma_l\,, \nonumber \\
S_i&=&-\frac{1}{2}\sigma_i+\frac{x_i}{r^2}(\vec{r},\vec{\sigma})
\,. \label{pwiy:eq19}
\end{eqnarray}

We see that $\vec{J}$ represents total angular momentum, 
$\vec{\Pi}$ a ``modified linear momentum'' and 
$\vec{S}$ a ``modified spin''. The algebra is isomorphic to a direct sum 
of the Euclidean Lie algebra $e(3)$ with the algebra $o(3)$
\begin{eqnarray}
\mathcal{L} \sim {e}(3)\oplus {o}(3) = \{\vec{J}-\vec{S}, \,\,\vec{\Pi}\}
\oplus \{\vec{S}\}
\,.\label{pwiy:eq20}
\end{eqnarray}

These generators satisfy the following commutation relations
\begin{eqnarray}
[J_i-S_i, S_j]=0\,, \qquad [\Pi_i, S_j]=0\,, \qquad [\Pi_i, \Pi_j]=0\,, 
\nonumber \\
\,[J_i-S_i, J_j-S_j]=i\epsilon_{ijk}(J_k-S_k)\,, 
\quad [J_i-S_i, \Pi_j]=i\epsilon_{ijk}\Pi_k
\,. \label{pwiy:neweq5}
\end{eqnarray}

It is interesting to note that the potentials in (\ref{pwiy:eq18}) are a 
purely quantum mechanical effect. Indeed if we reintroduce $\hbar$ into the 
Hamiltonian (\ref{pwiy:eq2}) and integral (\ref{pwiy:eq16}) it will figure 
significantly 
in the determining equations (\ref{pwiy:neweq1}), (\ref{pwiy:neweq3}) and 
(\ref{pwiy:neweq4}). The potentials in (\ref{pwiy:eq18}) are then modified
to
\begin{equation}
V_0=\frac{\hbar^2}{r^2}\,, \qquad V_1=\frac{\hbar}{r^2}
\,. \label{pwiy:neweq6}
\end{equation}
In the classical limit $\hbar \rightarrow 0$ both $V_0$ and
$V_1$ vanish.

Integrable and superintegrable quantum systems that have free
motion as their classical limits also exist in the case of scalar 
particles \cite{Gravel, Gravel2, Hietarinta} but they are
related to third- and higher-order integrals of motion.

\noindent {\bf (b) Spherical symmetry.}

For $V_1=V_1(r)$, $V_0=V_0(r)$ we obtain the well-known result that $H$
commutes with total angular momentum 
$\vec{J}=\vec{L}+\frac{1}{2}\vec{\sigma}$.

A full discussion will be presented elsewhere \cite{Winternitz3}.

\section{Conclusions}\label{pwiy:sec4}
We have shown that first-order integrability and superintegrability in the 
presence of spin--orbital interactions exist and are nontrivial. For $n=2$ 
the superintegrable potentials do not depend on $\hbar$ whereas for $n=3$ 
they vanish in the classical limit $\hbar \rightarrow 0$. Work is 
in progress on the search for superintegrable systems invariant under 
rotations and allowing second-order integrals of motion.

\section*{Acknowledgments}
We thank F. Tremblay who participated in the early stages of this project 
for helpful discussions. The work of P.W. was partly supported by a research 
grant from NSERC. \.{I}.Y. acknowledges a postdoctoral fellowship awarded 
by the Laboratory of Mathematical Physics of the CRM, 
Universit\'{e} de Montr\'{e}al.

\makeatletter
\renewcommand{\@biblabel}[1]{$^{#1}$}
\makeatother

\end{document}